\documentclass[journal,twoside,web]{ieeecolor}
\usepackage{generic}
\usepackage{cite}
\usepackage{amsmath,amssymb,amsfonts}
\usepackage{algorithmic}
\usepackage{graphicx}
\usepackage{algorithm,algorithmic}
\usepackage{hyperref}
\hypersetup{hidelinks=true}
\usepackage{textcomp}
\begin{document}
\title{Diffusion Probabilistic Multi-cue Level Set for Reducing Edge Uncertainty in Pancreas Segmentation}
\author{Yue Gou, Yuming Xing, Shengzhu Shi, and Zhichang Guo
\thanks{This work is supported in part by the National Natural Science Foundation of China (12171123, U21B2075, 12371419, 11971131, 11871133); in part by the Fundamental Research Funds for the Central Universities (2022FRFK060020); and in part by the Natural Sciences Foundation of Heilongjiang Province (ZD2022A001). (Corresponding authors: Shengzhu Shi.) }
\thanks{Yue Gou, Yuming Xing, Shengzhu Shi, and Zhichang Guo are with the Department of Computational Mathematics, School of Mathematics, Harbin Institute of Technology, Harbin 150001, China. (e-mail: 21b912028@stu.hit.edu.cn; xyuming@hit.edu.cn; mathssz@hit.edu.cn; mathgzc@hit.edu.cn).}
}

\maketitle

\begin{abstract}
Accurately segmenting the pancreas remains a huge challenge. Traditional methods encounter difficulties in semantic localization due to the small volume and distorted structure of the pancreas, while deep learning methods encounter challenges in obtaining accurate edges because of low contrast and organ overlapping. To overcome these issues, we propose a multi-cue level set method based on the diffusion probabilistic model, namely Diff-mcs. Our method adopts a coarse-to-fine segmentation strategy. We use the diffusion probabilistic model in the coarse segmentation stage, with the obtained probability distribution serving as both the initial localization and prior cues for the level set method. In the fine segmentation stage, we combine the prior cues with grayscale cues and texture cues to refine the edge by maximizing the difference between probability distributions of the cues inside and outside the level set curve. The method is validated on three public datasets and achieves state-of-the-art performance, which can obtain more accurate segmentation results with lower uncertainty segmentation edges. In addition, we conduct ablation studies and uncertainty analysis to verify that the diffusion probability model provides a more appropriate initialization for the level set method. Furthermore, when combined with multiple cues, the level set method can better obtain edges and improve the overall accuracy. Our code is available at https://github.com/GOUYUEE/Diff-mcs.
\end{abstract}

\begin{IEEEkeywords}
pancreas segmentation, diffusion probabilistic model, multi-cue segmentation, level set method, edge uncertainty. 
\end{IEEEkeywords}

\section{Introduction}
\label{sec:introduction}
\IEEEPARstart{P}{ancreatic} cancer is a highly incurable tumor, and early diagnosis is crucial for improving patient survival rates\cite{sung2021global}. Pancreas segmentation can assist doctors in calculating tumor volume and assessing its spread. Accurate segmentation of the pancreatic edges is in great demand in diagnosis\cite{khristenko2021preoperative}. However, the small volume, distorted structure, and low contrast between different tissues or organs often bring a lot of difficulties to the segmentation of the pancreas, leading to segmentation failure or edge blurring. In addition to the difficulties caused by the pancreas itself, pancreatic imaging data mainly comes from computed tomography(CT), and potential factors such as noise and artifacts generated by its imaging modality also pose challenges for pancreas segmentation tasks.

Traditional image segmentation methods, including region growing\cite{rusko2007fully}, level set\cite{cremers2007review}, graph cut\cite{peng20153d}, and variational models\cite{peng2014liver}, are difficult to cope with the aforementioned challenges. These methods often fail to accurately identify the location of the pancreas, leading to segmentation failure. Moreover, traditional methods require manual parameter tuning. For complex abdominal CT images, the position and shape of the pancreas can vary greatly, making it challenging to transfer the same parameters to other images. 

In contrast, learning-based approaches can segment the pancreas directly from a large amount of data by learning discriminative features through feature representations and segmentation patterns\cite{ghorpade2023automatic}. The early pancreas segmentation works mostly rely on performing volumetric multiple atlas registration and executing robust label fusion methods, which are referred to as top-down model fitting methods. For instance, Walz et al.\cite{wolz2013automated} developed a fully automatic multi-organ segmentation involving hierarchical atlas generation steps and refinement steps. The most suitable atlas is selected based on the global image appearance, aligned with the target image, and locally weighted on individual organs. Subsequently, segmentation refinement is performed using patch-based and graph-cut techniques that incorporate constraints related to local smoothness and higher-order spatial relationships. The Dice for the pancreas reaches 70\%.

In recent years, deep convolutional neural networks (CNNs) have been successfully applied to various tasks in medical image processing and have shown promising performance in pixel-level semantic segmentation. Some deep learning models have been proposed for pancreas segmentation\cite{roth2018spatial,chen2022target,yu2018recurrent,cai2019pancreas,li2023automatic,xu2022new}. To adapt to the complexity of pancreas, MobileNet-U-Net (MBU-Net)\cite{huang2022semantic} has been proposed, which combined the architectures of MobileNet-V2 and U-Net with repeated dilated convolutions for semantic pancreas segmentation. The Dice was improved to 82.87\% with fewer computational parameters. Typically, taking smaller input regions as a starting point leads to more accurate segmentation. A two-dimensional fixed-point model\cite{zhou2017fixed} of predicted pancreatic masks can narrow down the input region with a Dice of 83.18\%. On the other hand, Attention U-Net\cite{oktay2018attention} has been introduced, which combined the attention mechanism and U-Net architecture to selectively highlight regions relevant to pancreas segmentation in the input images. This attention mechanism was learned from the input image itself, allowing it to focus on information-rich regions and suppress irrelevant areas. The Dice reached 84\%, which is a 2.6\% improvement compared to the standalone U-Net architecture.

Generative Adversarial Networks (GANs) are a special type of CNN that optimize segmentation results through adversarial training between a generator and discriminator\cite{xue2018segan}. The generator network aims to produce realistic segmentation results, while the discriminator network is used to distinguish between real and generated segmentations, thereby driving improvements in the generator network. In particular, ADAU-Net was proposed by Li et al.\cite{li2022attention}, which introduced two-layer constraints into traditional networks through adversarial learning for pancreas segmentation, along with the inclusion of pyramid pooling modules. In this work, the average Dice for U-Net, Adversarial U-Net, and ADAU-Net were 77.37\%, 80.83\%, and 83.76\% respectively. In adversarial learning, attention blocks are integrated into multiple locations of the backbone paragrapher to enhance interdependence between pixels and achieve better segmentation performance.

In addition, the two-stage model from coarse to fine is proposed to use the results of coarse segmentation to guide the fine stage segmentation in a more complex manner, which can significantly improve the accuracy. Most researchers have applied deep learning to traditional image segmentation algorithms to extract high-level features and guide the segmentation process. For example, Zhang et al.\cite{zhang2021deep} obtained an initial delineation by combining multiple graphic registrations, which is integrated with precision delineation using 3D-CNN and level set method, achieving a Dice of 82\%. Hu et al.\cite{hu2020automatic} introduced DSD-ASPP-Net, which is a distance-aware model for learning the pancreatic position and probability maps. By combining with geodesic distance for detailed partitioning, the model can result in a Dice of 89.49\%. In addition, Tian et al.\cite{tian2023two} combined the 3D-Unet model with the level set method, which incorporates gradient direction priors and adaptive point parameters to improve edge delineation accuracy while reducing manual parameter tuning effects, achieving a Dice of 89.61\%. 

\begin{figure}[ht!]
\centering
\includegraphics[width=3.2in]{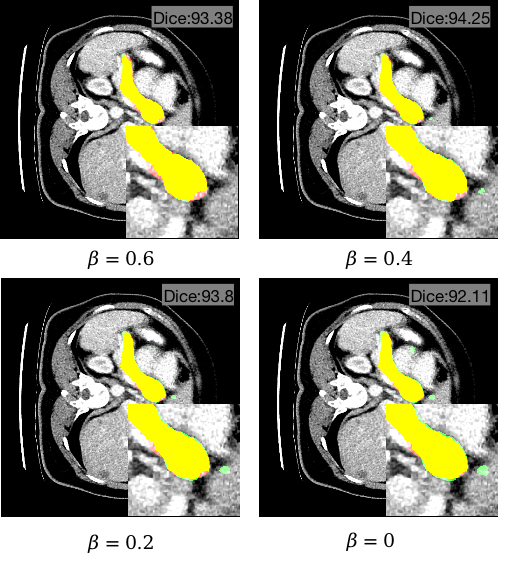}
\caption{According to different threshold values (\(\beta=0.6\), \(\beta=0.4\), \(\beta=0.2\), \(\beta=0\)) of the prior probability map, we obtain pancreas segmentation ground truth and inferred results respectively. Among these results, the red mask represents the pancreas segmentation ground truth, the green mask represents the inferred segmentation result, and the yellow area indicates accurately predicted regions.}
\label{beta}
\end{figure}
The two-stage learning framework has made a significant improvement over single-stage deep models. The main contribution lies in learning a prior probability map in the coarse stage and subsequently refining the edges and details in the fine stage. However, learning the prior probability map through end-to-end neural networks or GANs can be inappropriate. The output of end-to-end neural networks is not an approximation of the actual 'probability' of whether it is the target organ\cite{2020Gradient}. The output of GANs depends on the Nash equilibrium, which is neither the actual 'probability' nor unique. In comparison, probability models, such as the diffusion probabilistic model, can better serve for prior probability distribution estimation. More details will be discussed in the Motivation section.

Therefore, we propose a multi-cue level set method based on the diffusion probabilistic model(Diff-mcs). The diffusion probabilistic model estimates the prior probability distribution of the pancreas, and then the level set method incorporates grayscale cues, texture cues, and the prior cues to obtain fine segmentation results. The well-designed prior probability distribution serves two purposes in the level set method: firstly, it determines the initial position of the pancreas in a probabilistic manner without the need for parameter tuning, thus providing the initial contour of the level set. Secondly, the prior probability distribution, serving as a cue, offers stronger guidance to the curve evolution process and provides the level set method with a more reasonable direction. However, the probability model exhibits lower confidence at the pancreatic edge, often resulting in heightened uncertainty in edge segmentation due to the difficulties in obtaining adaptive thresholds. Luckily, the level set method can precisely serve as an adaptive threshold for confidence and provide more accurate and robust edge segmentation. 

The main contributions are as follows:

\subsubsection{}
We propose a coarse-to-fine segmentation strategy that utilizes the diffusion probabilistic model to learn the prior probability distribution in the coarse stage and applies a level set method to refine the edge in the fine stage, termed by Diff-mcs. The proposed framework achieves SOTA performance on three public datasets for pancreas segmentation.

\subsubsection{}
In the fine stage, we propose a multi-cue level set approach for pancreas segmentation, which not only includes regularization term and length term but also combines the prior probability distribution as prior cues, grayscale cues from the original image, and texture cues based on the nonlinear structural tensors. The inclusion of multiple cues allows for curve evolution to achieve finer segmentation, particularly on the edge of the pancreas.

\subsubsection{}
We conduct ablation studies and uncertainty
analysis to verify that the diffusion probabilistic model provides a more accurate estimation of the prior probability distribution compared to GANs. Furthermore, the multi-cue level set method precisely functions as an adaptive threshold for the diffusion probabilistic model. The diffusion probabilistic model integrates seamlessly with the level set method.

The remaining parts of this article are organized as follows: Section II introduces the motivation from the perspective of adaptive thresholding. The Diff-mcs method and technical details are described in Section III. Section IV presents the experimental setup, results, and analysis. Finally, Section V provides the conclusion of the entire paper.

\section{Motivation}
Traditional pancreas segmentation methods exhibit limited robustness against image texture variations and are sensitive to the choice of initial contours. Moreover, since the pancreas only occupies a small part of the entire CT image, the data imbalance can cause the models to tend towards more negative samples and potentially missing pancreatic regions. Deep learning methods can learn the complicated textures and avoid the choice of initial contours. However, the complex morphology and low contrast of the pancreas result in suboptimal segmentation in edges. But the problem can be well-solved with traditional methods. To this end, we come up with a coarse-to-fine learning framework that applies deep learning models to learn a prior probability distribution of the pancreas and further refines the edge areas with traditional methods.

In the selection of the deep model for the coarse stage, end-to-end neural networks like U-Net often tend to assign excessively high confidence to both positive and negative samples. Their output can function as the initial contour or cues for the upcoming fine stage, but they don't serve as good estimators of the prior probability distributions because they aren't actual 'probability'. Another option is GANs, however, they suffer from an unstable training process. Their un-unique equilibriums also lead to incorrect prior probability distribution. A more suitable option is probability models, such as the diffusion probabilistic model\cite{ho2020denoising} which can be employed to estimate the prior probability distribution by inferring the model through Monte Carlo. In particular, Amit et al.\cite{amit2021segdiff} combined the diffusion probabilistic model with the U-Net architecture, achieving segmentation results guided by the original image through end-to-end learning. 

Thus, we select the diffusion probabilistic model for the coarse stage. The diffusion probabilistic model generates data by gradually adding Gaussian noise, which is reversible at each step. As a Markov process, it has a better interpretation and can be directly designed as the prior probability distribution. Specifically, we employ MedSegDiff\cite{wu2022medsegdiff}, which has shown promising results across various types of medical images. To reduce the influence of high-frequency noise in medical images, MedSegDiff introduced a Feature Frequency Parser specifically designed for medical image segmentation. In particular, dynamic conditional encoding was proposed to establish state-adaptive conditions for each sampling step, which reduced the dependence of segmentation results on previous diffusion processes. 

Although diffusion probabilistic models show promising performance in medical image segmentation, they exhibit lower confidence at edges because it is difficult to determine an adaptive threshold, resulting in increased uncertainty in edge segmentation. In Fig.~\ref{beta}, we analyze the prediction of MedSegDiff using different thresholds (0.6, 0.4, 0.2, 0). It can be observed that, when all the relevant information is considered, the pancreas is under-segmented around the edge. At a threshold of \(\beta=0.4\), there is an improvement in pancreas segmentation and an increase in the Dice score, making it the best result among the four thresholds tested. At a threshold of \(\beta=0.6\), there are no more artifacts present in the image; however, over-segmentation occurs and leads to a decrease in Dice. It can be seen that there are differences in segmentation results obtained from the diffusion probabilistic model at different thresholds, making it difficult to find a correct threshold for determining the edges of segmented results.

\begin{figure}[ht!]
\centering
\includegraphics[width=3.8in]{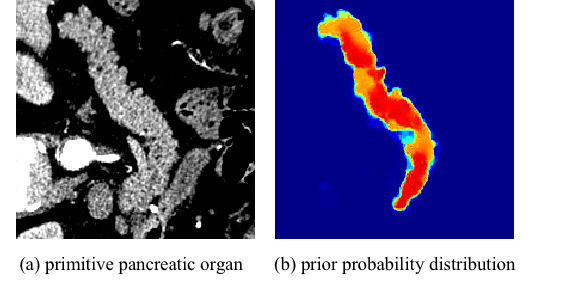}
\caption{Based on the diffusion probabilistic model, the pancreas segmentation results show that areas in the probability map closer to red indicate higher confidence of belonging to the pancreas, while areas closer to blue indicate lower confidence.}
\label{edge}
\end{figure}

In addition, we visualize the probability map obtained by the diffusion probabilistic model. In Fig.~\ref{edge}, we can observe that the probability map favors higher probabilities in the middle and lower probabilities at the edges\cite{gregor2016towards}. The edge ground truth passes through varying confidence thresholds; thus, a constant threshold is inappropriate. 

Therefore, in the fine segmentation stage, we hope to use the multi-cue level set method to refine the edge, or in other words to act as the adaptive threshold. The prior probability map can also guide the evolution of curves during the segmentation process and enable flexible handling of objects with complex shapes. Additionally, it can provide extra edge constraints to help the level set method capture and maintain edge accuracy more effectively.

\begin{figure*}
  \includegraphics[width=7.2in]{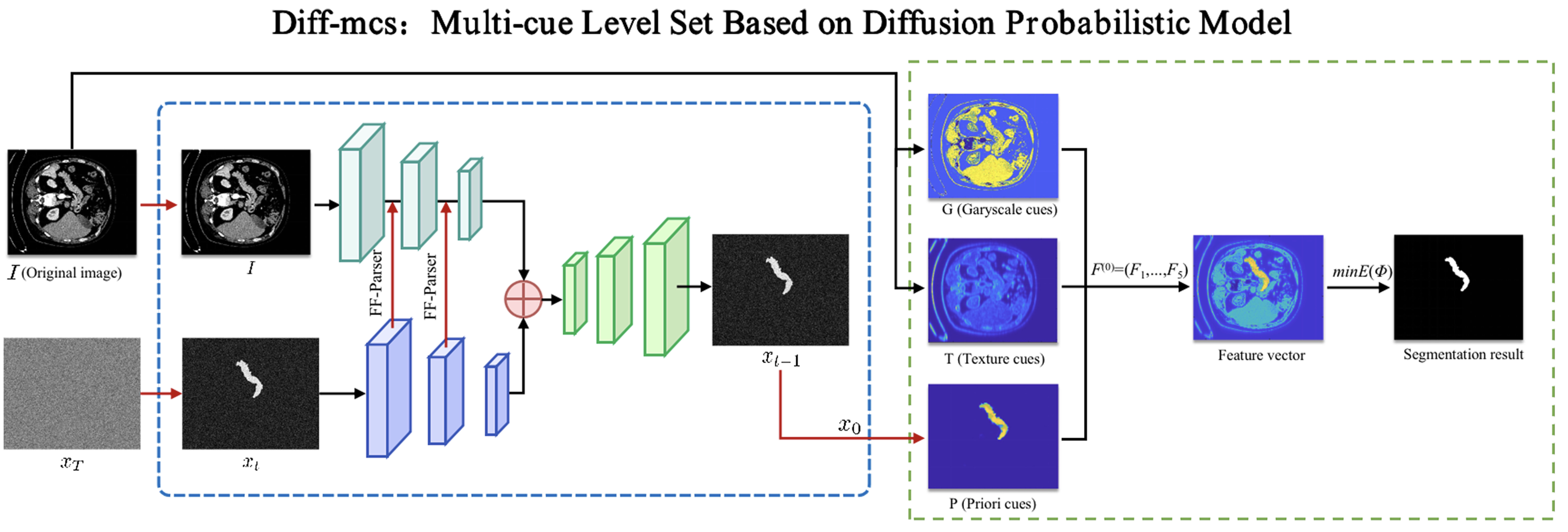}
  \caption{The multi-cue level set method is guided by a diffusion probabilistic model. Where the blue box is the process of the diffusion probabilistic model, and the green box represents the multi-cue level set method. In the blue box, we implement it with a ResNet encoder following a U-Net decoder. The encoder consists of a set of condition encoders and segmentation encoders with an FF-parser on the feature fusion path to constrain noise and connection. In the green box, The multi-cue level set method combines grayscale cues, texture cues, and prior cues to segment the pancreas by determining the initial level set positions using the prior probability distributions obtained from the network.}
  \label{pancreatic segmentation.}
\end{figure*}

In the multi-cue segmentation methods, Feng et al.\cite{feng2016interactive} proposed a novel interactive segmentation algorithm by combining color, depth, normals and other cues under the graph cut framework. The model used a cue to determine the segmentation label based on intuition and automatically selected the cue used on each pixel, which can not only obtain the segmentation mask but also generate the cue label map. In the framework of the level set method, Thomas et al.\cite{brox2010colour} utilized nonlinear structural tensors to extract and enhance image features and combined different cues from color, texture, and motion for image segmentation.

The multi-cue level set method aims to maximize the difference between probability distributions of cues inside and outside curve to achieve image segmentation by adjusting the level set curve based on gradient direction and intensity. Inspired by this, we naturally consider incorporating the probability distribution obtained from diffusion probabilistic models as prior cues into the multi-cue level set method. The prior cues are used to obtain a reasonable initial contour and assist the level set method in fine segmentation. Moreover, edges encode shape information while textures determine region information. We utilize nonlinear structure tensors to extract and enhance image features as texture cues. In conclusion, combining grayscale cues, texture cues, and prior cues, the proposed model can enrich feature representation. Therefore, the diffusion probabilistic model integrates with the level set method that shows the potential to improve image segmentation.

\begin{figure*}
  \centering
  \includegraphics[width=7in, trim=-0.4in 0 0 0, clip]{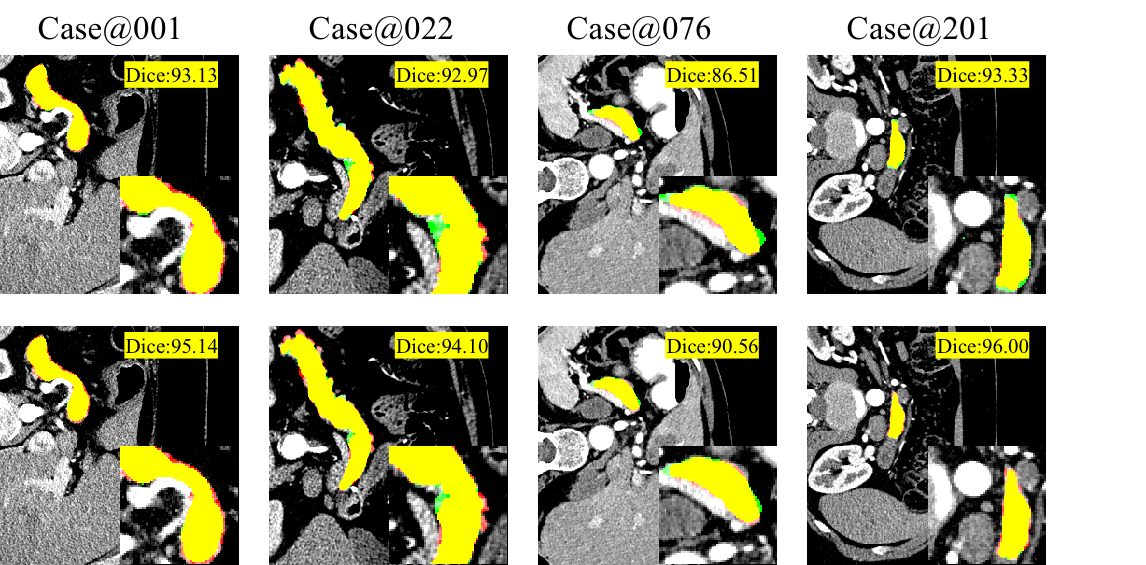}
  \caption{The first row shows the segmentation results of the diffusion probabilistic model, and the second row shows the segmentation results of Diff-mcs. The red contour is the gold standard, and the green contour is the predicted segmentation results.}
  \label{abk}
\end{figure*}

\section{Methodology}
\label{sec:Methodology}
In this section, we describe the details of Diff-mcs. Diff-mcs aims to utilize the prior probability distribution obtained by the diffusion probabilistic model to assist in fine segmentation. As shown in Fig.~\ref{pancreatic segmentation.}, firstly, a segmentation network based on diffusion probabilistic model\cite{wu2022medsegdiff} is employed, which is implemented with a ResNet encoder following a U-Net decoder to obtain a coarse pancreatic probability map. The probability map is used for locating the initial position of the pancreas. Additionally, the prior probability distribution serves as one of the cues for the level set method. Then, during the fine segmentation stage, a multi-cue level set incorporating grayscale cues extracted from original images, texture cues based on nonlinear structure tensors, and prior cues derived from the probability map is employed to enhance model performance.
\subsection{Diffusion Probabilistic Model}
Our first stage refers to MedSegDiff and adopts a segmentation structure based on the diffusion probabilistic model. This structure integrates the ideas of the diffusion process and probability model, gradually generating samples for segmentation through multiple steps of diffusion operations. The core idea is to utilize the diffusion process to transform a simple initial distribution into the target distribution, including both forward and reverse processes.
\subsubsection{Forward Process}
The diffusion probabilistic model is a generative model parameterized by a Markov chain, with the forward process defined as follows:
\begin{equation}
q(x_{1:T} | x_0) = \prod_{t=1}^{T} q(x_t | x_{t-1}),
\end{equation}
where \(T\) represents the number of steps in the diffusion model, \(x_{1:T}\) are the latent variables, and \(x_0\) is real segmentation results from the data. 
According to the Markov process, gradually adding Gaussian noise into the data can be described as:
\begin{equation}
q(x_t | x_{t-1}) = \mathcal{N}(x_t; \sqrt{1 - \beta_t} x_{t-1}, \beta_t I_{n \times n}),
\end{equation}
where the sequence of noise coefficients \(\beta_{1:T} \) are constants that define the linear growth of adding noise and \(I_{n \times n}\) is the identity matrix of size \(n\). 

The forward process at any given time \(t\) can be defined as:
\begin{equation}
q(x_t | x_0) = \mathcal{N}\left(x_t; \sqrt{\bar{\alpha}_t} x_0, (1 - \bar{\alpha}_t)I_{n \times n}\right),
\end{equation}
where \[\alpha_t := 1 - \beta_t; \quad \bar{\alpha}_t :=\prod_{s=0}^t \alpha_s,\]
which can be reparametrized to:
\begin{equation}
x_t = \sqrt{\bar{\alpha}_t} x_0 + \sqrt{(1 - \bar{\alpha}_t)}\varepsilon, \quad \varepsilon \sim \mathcal{N}(0, I_{n \times n}).
\end{equation}
\subsubsection{Reverse Process}
Calculating \(q(x_{t-1}|x_t, x_0)\) using Bayes' theorem, one obtains:
\begin{equation}
\begin{aligned}
q(x_{t-1}|x_t, x_0) & = \frac{q(x_{t}|x_{t-1})q(x_{t-1}|x_0)}{q(x_{t}|x_0)}\\
& = \mathcal{N}(x_{t-1}; \tilde{\mu}(x_t, x_0), \tilde{\beta}_t I_{n \times n}),
\end{aligned}
\end{equation}
where
\[
\tilde{\mu}(x_t, x_0) = \frac{\sqrt{\bar\alpha_{t-1}}\beta_t}{1 - \bar\alpha_{t}}x_0 + \frac{\sqrt{\alpha_t}(1 - \bar\alpha_{t-1})}{1 - \bar\alpha_{t}}x_t,
\]
\[
\tilde{\beta} = \frac{1 - \bar\alpha_{t-1}}{1 - \bar\alpha_{t}}\beta_t.
\]

The probability distribution of the reverse process can be parameterized by \(\theta\) as:
\begin{equation}
p_{\theta}(x_{0:T-1}|x_T) = \prod_{t=1}^{T} p_{\theta}(x_{t-1}|x_t).
\end{equation}

The reverse process transforms the latent variable distribution \(p_{\theta}(x_T)\) to the data distribution \(p_{\theta}(x_0)\). The reverse process can be described as:
\begin{equation}
p_{\theta}(x_{t-1}|x_t) = \mathcal{N}(x_{t-1}; \mu_{\theta}(x_t, t), \Sigma_{\theta}(x_t, t)),
\end{equation}
which follows a Gaussian distribution \(\mathcal{N}\) with mean \(\mu_{\theta}(x_t, t)\) and variance \(\Sigma_{\theta}(x_t, t)\).

The noise \(\varepsilon_{\theta}\) predicted by the neural network \(\mu_{\theta}\) can be parameterized as:
\begin{equation}
\mu_{\theta}(x_t, t) = \frac{1}{\sqrt{\alpha_t}} (x_t - \frac{1-\alpha_{t}}{\sqrt{1-\bar\alpha_{t}}} \varepsilon_{\theta}(x_t, t)).
\end{equation}

For inference, letting $\sigma_t^2=\tilde\beta$, we reparameterize the reverse process as:
\begin{equation}
x_{t-1}=\frac{1}{\sqrt{\alpha_t}}\left(x_t-\frac{1-\alpha_t}{\sqrt{1-\bar{\alpha}_t}} \varepsilon_\theta\left(x_t, t\right)\right)+\sigma_t z, z \sim \mathcal{N}(0, I_{n \times n}).
\end{equation}

\begin{figure*}
  \centering
  \includegraphics[width=7in, trim=-0.8in 0 0 0, clip]{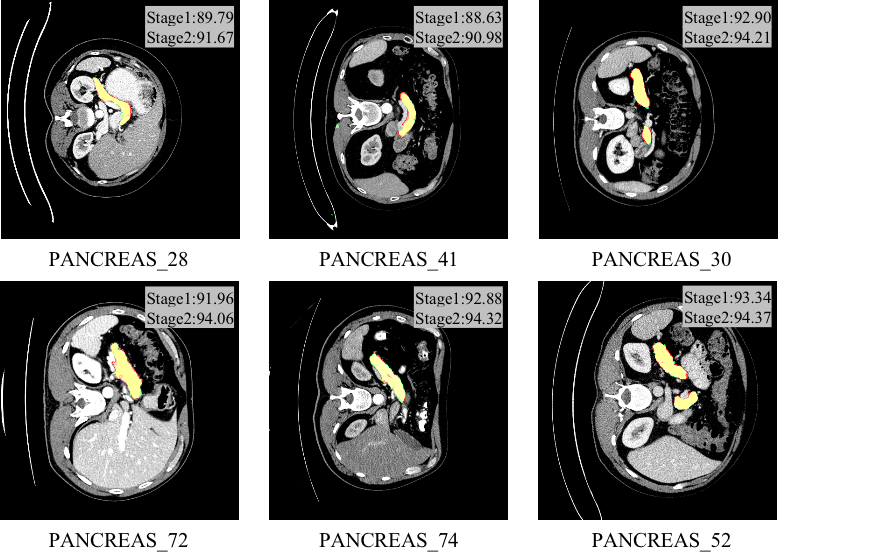}
  \caption{The results on the NIH dataset, the red edge line represents the real mask, the green mask represents the first stage coarse segmentation results, the light red mask represents the Diff-mcs results, the overlap is indicated by yellow, and the upper right corner is the Dice of the two stages respectively.}
  \label{nih}
\end{figure*}
\subsubsection{Diffusion Training Objective}
The training objective utilizes KL divergence by minimizing the negative log-likelihood of the variational lower bound:
\begin{equation}
\begin{aligned}
\mathbb{E}\left[-\log p_\theta\left(\mathrm{x}_0\right)\right] & \leq \mathbb{E}_q[\underbrace{D_{\mathrm{KL}}\left(q\left(x_T \mid x_0\right) \| p\left(x_T\right)\right)}_{L_T} \\
& +\sum_{t>1} \underbrace{D_{\mathrm{KL}}\left(q\left(x_{t-1} \mid x_t, x_0\right) \| p_\theta\left(x_{t-1} \mid x_t\right)\right)}_{L_{t-1}} \\
& \underbrace{\left.-\log p_\theta\left(x_0 \mid x_1\right)\right].}_{L_0}
\end{aligned}
\end{equation}

With the help of variational inference\cite{ho2020denoising}, we minimize the term:
\begin{equation}
\mathbb{E}_{x_0, \varepsilon, t}\left[\left\|\varepsilon-\varepsilon_\theta\left(\sqrt{\bar{\alpha}_t} x_0+\sqrt{1-\bar{\alpha}_t} \varepsilon, t\right)\right\|^2\right],
\end{equation}
where $\varepsilon \sim N\left(0, I_{n \times n}\right)$.
\subsubsection{Segmentation Based on Diffusion Probabilistic Model}
In the diffusion probabilistic model of the segmentation process, we provide an additional input of the original image \(I\) to guide the generation of segmentation results by the diffusion probabilistic model. \(\varepsilon_{\theta}\) is typically a U-Net represented as follows:
\begin{equation}
\varepsilon_{\theta}(x_t, I, t) = D\left(E_A\left(E_B(x_t,t) + E_C(I,t), t\right), t\right).
\end{equation}

In this architecture, the decoder \(D\) of the U-Net is conventional, while its encoder is decomposed into three networks: \(E_A\), \(E_B\), and \(E_C\). \(E_C\) represents conditional feature embedding, which embeds the original image; \(E_B\) represents feature embedding of the segmentation map for the current step. They are connected to the FF-parser in a path for feature integration, which restricts components related to noise in \(x_t\) features. The encoders consist of three convolutional stages. The residual blocks for each stage follow the structure of ResNet34, comprising two convolutional blocks with group-norm and SiLU\cite{elfwing2018sigmoid} active layer, as well as a convolutional layer. These two processed inputs have the same spatial dimensions and channel numbers and are summed up as signals. The summation is then passed onto the remaining part of U-Net's encoder \(E_A\) and sent to U-Net's decoder \(D\) for reconstruction. The time step \(t\) is integrated with embeddings.
\begin{figure*}
  \centering
  \includegraphics[width=7in]{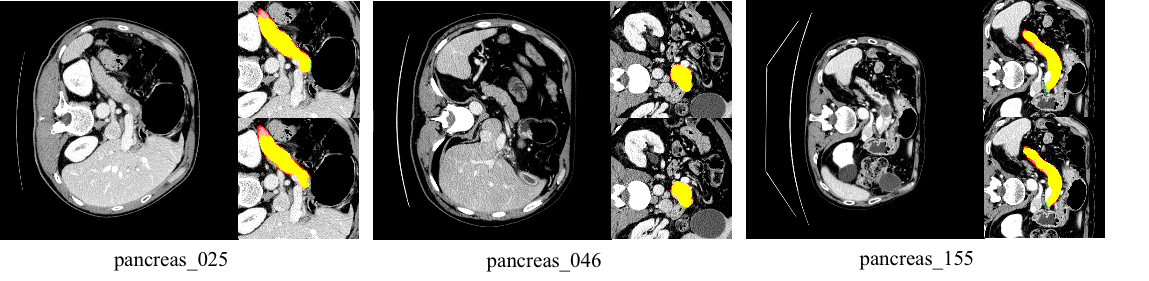}
  \caption{The results on the MSD dataset, the large graph is the original graph, and the two attached subgraphs, the upper one is the resulting graph of coarse segmentation, and the lower one is the resulting graph of fine segmentation.}
  \label{msd}
\end{figure*}

\subsection{Multi-cue Level Set Method}
Although the diffusion probabilistic model can locate the position of the pancreas, it's unable to describe its precise edges. Therefore, we introduce a multi-cue level set method that incorporates prior probability distributions.

Let $\Omega$ be a two-dimensional image domain, where each pixel can be represented as $(r, c)$ and the pixel value can be represented as $I(r, c)$. Our goal is to segment the image into two different regions $\Omega_1$ and $\Omega_2$ so that pixels within each region have similar characteristics or belong to the same object. To simplify the expression, in subsequent discussions, we will omit $c$ and use $I(r)$ to represent the pixel value of the image.

For the level set method, a level set function $\phi(r)$ is used to describe the edges or contours in the image. The function approaches zero when close to a boundary; positive values indicate being inside the boundary that $r \in \Omega_1$; negative values indicate being outside the boundary that $r \in \Omega_2$. The evolution process of the level set function $\phi(r)$ is described by the following evolution equation\cite{aubert2006mathematical}:
\begin{equation}
\begin{cases}
\frac{\partial \phi}{\partial t} = F_t|\nabla\phi|, \quad\quad (r, t) \in \Omega \times (0, T], \\
\phi(r, 0) = \phi_0(r), \quad\quad\quad r \in \Omega,
\end{cases}
\end{equation}
where $\nabla\phi$ represents the spatial gradient of $\phi$ and $|\nabla\phi|$ represents its magnitude, and $\phi_0(r)$ is the initial value. $F_t$ is a driving force function related to the normal component of the velocity, which does not change sign during the evolution and the orientation depends on the type of evolution. This driving force function is usually determined based on image features and segmentation task requirements.

Our approach combines grayscale cues, texture cues, and prior cues into five feature channels: one for image grayscale values, three for different components of spatial structural characteristics, and the other one for the prior probability distributions.
The introduced feature vectors are based on classical linear structure tensors:
\[
J_{\rho} = K_{\rho} * \left( \nabla I  \nabla I^{T} \right) = \left( \begin{array}{cc} K_{\rho} * I_{r}^{2} & K_{\rho} * \left(I_{r}I_{c}\right) \\ K_{\rho} * \left(I_{c}I_{r}\right) & K_{\rho} * I_{c}^{2} \end{array} \right),
\]
where \(K_{\rho}\) is a Gaussian kernel, with a standard deviation parameter \(\rho\) that determines the scale of the neighborhood used for analysis, and \(I\) is the grayscale image.

In order to solve the problem that edge attenuation caused by Gaussian smoothing makes the results inaccurate at data discontinuities, we use nonlinear diffusion instead of Gaussian smoothing\cite{weickert1997review} as follows:
\begin{equation}
\label{diffusion}
\frac{\partial F_T}{\partial t} = div\left(g\left(|\nabla F_T|^2\right) \nabla F_T\right)  
\end{equation}
with $F_T(t=0)$ being the initial image and $g$ is a decreasing diffusivity function as follows:
\begin{equation}
g(|\nabla F_T|) = \frac{1}{({|\nabla F_T|^2 + \tau^2})^{\frac{p}{2}}}
\end{equation}
Here $\tau$ is used to avoid the singularity of the diffusion function at zero, while the value of $p$ determines various characteristics of the diffusion function, such as its ability to preserve and enhance edges as well as the smoothness exhibited by the resulting image.
The first step is to normalize the feature vector: \(F^{(0)} = (F_1, \ldots, F_5)\). The initial texture features \(F_{T}^{(0)} := ((I_{0})_{r}^{2};(I_{0})_{c}^{2};2(I_{0})_{r}(I_{0})_{c})\) with \(I_{0}\) being the image gray value. The AOS algorithm\cite{weickert1998anisotropic} can be utilized for the implementation of the above diﬀusion equation\eqref{diffusion}. Next, we can get the normalised texture feature as follows\cite{brox2010colour}: \(\widetilde{F_{T}} = \left[ \begin{array}{cc} \widetilde{F_{T_{11}}} & \widetilde{F_{T_{12}}} \\ \widetilde{F_{T_{21}}} & \widetilde{F_{T_{22}}} \end{array} \right]=\sqrt{F_T} = T(\sqrt{\lambda_i})T^\intercal\). So the ﬁnal feature vector are \(F = (F_1, \ldots, F_5) = (I_{0}, 2\widetilde{F_{T_{11}}}, 2\widetilde{F_{T_{22}}}, 4\widetilde{F_{T_{12}}}, P)\) with \(P\) being the prior cues.

\begin{algorithm*}[t]
	\caption{Multi-cue Level Set Method.}
	\label{alg:alg1}
    \textbf{Input }{the original image $I_{0}$, prior probability distribution $P$, the step number $N$, parameters $\mu, \nu, \sigma, p, \tau, \omega_{j}$, the step size $\eta$}
	
    \textbf{Output }{$\phi^{n+1}$}
	
	Locate the initial contour curve $\phi^0=\phi_{0}$.
	
	\textbf{For }$n = 1,2,\ldots,N${
		
           \hspace{0.6cm}Compute the initial value $F_{T}^{(0)} := ((I_{0})_{r}^{2};(I_{0})_{c}^{2};2(I_{0})_{r}(I_{0})_{c})$.
		
		   \hspace{0.6cm}Using AOS algorithm solve \eqref{diffusion}.
		
		   \hspace{0.6cm}Compute the normalized feature vector $F := (F_{1},\ldots,F_{5}) = (I_{0}, 2\widetilde{F_{T_{11}}}, 2\widetilde{F_{T_{22}}}, 4\widetilde{F_{T_{12}}}, P)$.
		
		   \hspace{0.6cm}Compute the probability density functions $p_{ij}$ given by \eqref{p}, $(i=1,2, j=1,2,3,4,5)$.
		
		   \hspace{0.6cm}Update $\phi^{n}$ through the iterative equation \eqref{lsf}.
     
     \textbf{end}
	}
\end{algorithm*}

Then, we define each region as \(i\) and each feature channel as \(j\), with mean \(\mu_{ij}\) and standard deviation \(\sigma_{ij}\) describing the distribution of each region and feature channel, disregarding inter-channel correlations. \(p_{ij}\) represents the probability functions associated with each feature \(F_j\). we can express the fitting term as a problem of minimizing energy:
\begin{equation}
\begin{aligned}
E(\Omega_i, p_{ij}) &= -\sum_{j=1}^{N} \int_{\Omega_1} \log p_{1j}(F_j(r)) \, dr \\
& + \sum_{j=1}^{N} \int_{\Omega_2} \log p_{2j}(F_j(r)) \, dr.
\end{aligned}
\end{equation}
where
\[
\mu_{ij} = \frac{\int_{\Omega} F_j(r)\mathcal{X}_i(r) \, dr}{\int_{\Omega}\mathcal{X}_i(r) \, dr},
\]
\begin{equation}
\label{p}
\sigma_{ij} = \sqrt{\frac{\int_{\Omega} \left(F_j(r) - \mu_{ij}\right)^2\mathcal{X}_i(r) \, dr}{\int_{\Omega}\mathcal{X}_i(r) \, dr}},
\end{equation}
Here, \(\mathcal{X}_i\) are the indicator functions that map \(s\) to \(\mathcal{X}_i(s)\).

In addition, the Heaviside function is approximated by a smooth function \(H_{\tau}\) defined as:
\[
H_{\tau}(r) = \frac{1}{2}\left(1 + \frac{2}{\pi}\arctan\left(\frac{r}{\tau}\right)\right).
\]

Similarly, we can use the level set method to solve this energy minimization problem. Let \(\mathcal{X}_1(s) = H_{\tau}(s)\) and \(\mathcal{X}_2(s) = 1 - H_{\tau}(s)\). The problem then becomes:
\begin{equation}
E(\phi, p_{ij}) = -\sum_{i=1}^{2} \sum_{j=1}^{N} \int_{\Omega} \log p_{ij}(F_j) \, \mathcal{X}_i(\phi(r)) \, dr.
\end{equation}

The objective is to find the optimal level set function \(\phi\) and probabilities \(p_{ij}\) that minimize the energy functional \(E(\phi, p_{ij})\).

Thus, we propose a multi-cue level set method by combining regularization term\cite{li2010distance} and length term as follows:
\begin{equation}
\begin{aligned}
\label{E}
E(\phi, p_{ij}) &= \frac{1}{2} \nu \int_{\Omega} (|\nabla \phi| - 1)^2 \, dr + \mu \int_{\Omega} g_I \delta(\phi)|\nabla \phi| \, dr\\
& + \omega_{j} \sum_{i=1}^{2} \sum_{j=1}^{N} \int_{\Omega} \log p_{ij}(F_j) \, \mathcal{X}_i(\phi(r)) \, dr,
\end{aligned}
\end{equation}
Here, the derivative of \(H_{\tau}\) is 
\[
\delta_{\tau}(r) = H_{\tau}'(r) = \frac{1}{\pi}\frac{\tau}{r^2 + \tau^2},
\]
In addition, we define an edge indicator function \(g_\phi\) by
\[
g_I = \frac{1}{1+\lvert \mathcal{}K_{\rho} * I(r) \rvert}.
\]

We can obtain the level set function evolution equation by the gradient flow method:
\begin{equation}
\label{gf}
\begin{cases}
\begin{aligned}
\frac{\partial \phi}{\partial t} &= \nu\left(\nabla^2 \phi - \text{div}\left(\frac{\nabla\phi}{|\nabla \phi|}\right)\right) + \mu \delta_\tau(\phi)\text{div}\left(g_I \frac{\nabla\phi}{|\nabla \phi|}\right)\\
&+ \omega_{j}\sum_{j=1}^{N}\log \frac{p_{1j}(F_j)}{p_{2j}(F_j)} H_{\tau}'(\phi), \quad\quad (r, t) \in \Omega \times (0, T],
\end{aligned}\\
\phi(r, 0) = \phi_0(r), \quad\quad\quad\quad\quad\quad\quad\quad\quad\quad r \in \Omega.
\end{cases}
\end{equation}

Among them, the first term of the \eqref{E} is the regularization term, which is used to maintain the regularity of the level set function. By introducing the distance regularization term, the deviation between the level set function and the signed distance function can be penalized, which promotes the level set function to maintain its regularity and prevents re-initialization. The second term is the length term. Parameterizing the level set function as a contour, ensures smoothness and continuity of the level set function during its evolution, thereby constraining its evolution.

\section{Experiments}
\subsection{Implementation Details}
\subsubsection{Dataset Description}
In this section, we will introduce three publicly available abdominal CT datasets that include the pancreas.

AbdomenCT-1K dataset\cite{ma2021abdomenct}: AbdomenCT-1K is a large-scale multi-organ dataset consisting of 1,112 abdominal CT scans, specifically designed for liver, kidney, spleen, and pancreas segmentation. This dataset integrates multiple publicly available datasets and provides comprehensive annotations for each CT scan. Currently, only the annotations for 1,000 cases have been made publicly available by the official source. We extracted the annotations specifically related to the pancreas.

NIH Pancreas-CT dataset\cite{roth2015deeporgan}: The NIH Pancreas dataset is a collection of data gathered by the Clinical Center of the National Institutes of Health in the United States. It consists of 82 sets of enhanced CT scans, with 17 being from healthy kidney donors and the remaining 65 from patients without abdominal lesions or pancreatic cancer. These CT scans have been manually annotated by medical students and verified and modified by experienced radiologists.

MSD Pancreas-CT dataset\cite{antonelli2022medical}: The MSD pancreas dataset consists of 281 annotated training data samples of the pancreas and tumors, as well as 139 test data samples, provided by the Memorial Sloan Kettering Cancer Center. These data are derived from patients who have undergone pancreatic tumor resection surgery, including various types of tumors. Each slice has been manually labeled by experts to identify the pancreatic parenchyma and lesions.

\subsubsection{Evaluation Metrics}
The symbol $PS$ represents the set of predicted segmentation pixels, while $GT$ represents the set of ground truth segmentation pixels. We define the symbol $|\cdot|$ to represent the cardinality of a set, corresponding to the number of pixels it contains.

Dice coefficient (Dice):
\[
\text{Dice} = \frac{2|PS \cap GT|}{|PS| + |GT|}.
\]
The Dice coefficient measures the overlap between the prediction and the ground truth segmentation results. The Dice coefficient ranges from 0 to 1, where a value closer to 1 indicates a higher overlap between the predicted and ground truth results.

Jaccard similarity coefficient (IoU):
\[
\text{IoU} = \frac{|PS \cap GT|}{|PS \cup GT|}.
\]
The Jaccard similarity coefficient measures the overlap between the predicted and ground truth segmentation results. The IoU ranges from 0 to 1, where a value closer to 1 indicates a higher overlap between the predicted and ground truth results.

\begin{table*}
\caption{five-fold cross-validation on the AbdomenCT-1K dataset}
\label{table1}
\setlength{\tabcolsep}{0pt}
\renewcommand{\arraystretch}{1.5}
\begin{tabular}{p{40pt}|p{80pt}|p{80pt}|p{80pt}|p{80pt}|p{80pt}|p{80pt}}
\hline
 & 
\hspace{1.1cm}Dice& 
\hspace{0.9cm}Jaccard&
\hspace{0.8cm}Accuracy&
\hspace{0.8cm}Precision&
\hspace{1cm}Recall&
\hspace{0.8cm}F1 score\\
\hline
\hspace{0.3cm}Stage1& 
\hspace{0.6cm}89.43$\pm$3.46 & 
\hspace{0.6cm}82.20$\pm$5.74 &
\hspace{0.6cm}99.75$\pm$7.55 & 
\hspace{0.6cm}95.88$\pm$2.24 & 
\hspace{0.6cm}85.38$\pm$6.48 & 
\hspace{0.6cm}90.34$\pm$3.46 \\
\hspace{0.3cm}Stage2 &
\hspace{0.6cm}92.30$\pm$4.69 &
\hspace{0.6cm}83.52$\pm$9.59 &
\hspace{0.6cm}99.78$\pm$2.45 & 
\hspace{0.6cm}95.10$\pm$7.55 &
\hspace{0.6cm}87.49$\pm$8.60 & 
\hspace{0.6cm}90,82$\pm$6.71\\
\hline
\end{tabular}
\end{table*}

\begin{table}
\caption{Comparison on the AbdomenCT-1K Dataset}
\label{tableabk}
\setlength{\tabcolsep}{3pt}
\renewcommand{\arraystretch}{1.2}
\begin{tabular}{p{80pt}|p{70pt}|p{70pt}}
\hline
\hspace{1cm}Method &
\hspace{0.5cm}Stage1(Dice)&
\hspace{0.5cm}Stage2(Dice)
\\
\hline
\hspace{0.5cm}Xu et al.\cite{xu2022new} & 
\hspace{1.2cm}-&
\hspace{0.5cm}74.60$\pm$14.90\\
\hspace{0.5cm}Ma et al.\cite{ma2021abdomenct} & 
\hspace{1.2cm}-&
\hspace{0.5cm}82.50$\pm$19.00 \\
\hspace{0.5cm}Li et al.\cite{li2023automatic} & 
\hspace{1.2cm}-&
\hspace{0.5cm}86.20$\pm$6.90 \\
\hspace{0.5cm}Tian et al.\cite{tian2023two} & 
\hspace{0.5cm}87.56$\pm$6.58 & 
\hspace{0.5cm}89.61$\pm$5.70  \\
\hspace{0.5cm}The proposed & 
\hspace{0.5cm}\textbf{89.43$\pm$3.46} &
\hspace{0.5cm}\textbf{92.30$\pm$4.69}  \\
\hline
\end{tabular}
\label{tableabk}
\end{table}

\subsubsection{Experimental Setting}
For image preprocessing, all our input 2D CT image slices have a size of \(512\times512\). Experimental results indicate that pancreas densities typically fall within the range of 40 to 70 Hounsfield Units (HU). Therefore, by setting the window width to 250 and window level to 50, we effectively adjust the intensity values of all pixels to [-75,175] HU to better visualize the pancreas.

 In the coarse segmentation stage, we train our model using an RTX 3090 GPU with an initial learning rate of 0.0001, diffusion steps of 1,000 iterations, and a batch size of 2 samples. We employ five-fold cross-validation on the AbdomenCT-1K dataset, in which we extract 2,000 2D slices from the first 400 3D cases as our training data. During the coarse segmentation process, each training iteration requires 400,000 to 450,000 iterations. For the testing phase, we use an ensemble of 10 models. Similarly, for the NIH datasets and MSD datasets, we select 1,800 2D slices, with 1,500 slices used for training and 300 slices used for testing.

In the fine segmentation stage, the Algorithm~\ref{alg:alg1} used is as follows, all the partial derivatives $\frac{\partial \phi}{\partial r}$ and $\frac{\partial \phi}{\partial c}$ in \eqref{gf} can be
simply discretized as central finite differences, and the temporal derivative is discretized as a forward difference. An iteration scheme is then obtained by discretizing the \eqref{gf} as follows: 
\begin{equation}
\begin{aligned}
\phi^{n+1}&=\phi^{n}+\eta\cdot(\nu(\nabla^2 \phi^{n} - \text{div}(\frac{\nabla\phi^{n}}{|\nabla \phi^{n}|}))\\
&+ \mu\delta_\tau(\phi^{n})\text{div}(g_I \frac{\nabla\phi^{n}}{|\nabla \phi^{n}|}) + \omega_{j}\sum_{j=1}^{N}\log \frac{p_{1j}(F_j)}{p_{2j}(F_j)} H_{\tau}'(\phi^{n})).   
\end{aligned}
\label{lsf}
\end{equation}
The parameter settings are as follows: \(\eta = 0.1\), \(\nu = 0.01\), \(\tau = 0.01\), \(p = 1.6\), \(\sigma = 3.0\)\ and \(\mu = 0.001 \times 255 \times 255\).

To demonstrate the role of the prior probability distribution in the model, we conducte ablation studies on the AbdomenCT-1K dataset for verification. Our ablation studies mainly focus on the proposed multi-cue level set method. From the 2,000 results of the first-stage five-fold cross-validation, we randomly select 400 cases as the dataset for the ablation studies.

\subsection{Experimental Results}
\subsubsection{Segmentation Results on AbdomenCT-1K Dataset}

In this section, we employ a five-fold cross-validation to evaluate the performance of Diff-mcs on the AbdomenCT-1K dataset.

As shown in Table \ref{tableabk}, the coarse segmentation based on the diffusion probabilistic model achieves a Dice score of 89.43$\pm$3.46 on the AbdomenCT-1K dataset. The Dice score of the coarse segmentation increased by 1.87 compared to the coarse stage of \cite{tian2023two}, whose coarse segmentation model is based on 3D U-Net. Compared to single-stage models, the coarse stage of Diff-mcs already has better performance. The satisfactory coarse segmentation performance of Diff-mcs shows that the diffusion probabilistic model can provide accurate initial localization and prior probability distribution for the fine stage. 

In the fine segmentation stage, Diff-mcs achieve an average Dice score of 92.30$\pm$4.69. Compared to the coarse stage, the fine stage provides an improvement of 2.87 in the Dice score. The performance boost in the fine stage enlarges the advantage to 3.23 in the Dice score compared to \cite{tian2023two}. The initial localizations are similar for two-stage models; however, the diffusion probabilistic model is more appropriate for probability distribution estimation. As shown in Table \ref{table1}, the fine stage has significant improvement to the coarse stage on Dice, Jaccard, accuracy, recall, and F1 score. It results in a slice decrease in precision since it calculates a more cautious and accurate edge.

In Fig.~\ref{abk}, we can see that our multi-cue level set method performs better in handling edges and calibrating rough segmentation edges. Both Case@001 and Case@022 improve the under-segmentation issue in coarse segmentation, as indicated by the reduced green areas. Additionally, Case@022, as shown in Fig.~\ref{edge}, demonstrates significant improvement in regions with low probabilities after being segmented using the multi-cue level set method. In the last two cases, both the head and tail of the pancreas are under-segmented in coarse segmentation. However, after applying the multi-cue level set method, these two regions show significant improvement. This is because the level set method utilizes curve evolution for segmentation, which has advantages in maintaining continuity and smoothness. The curves are attracted to edges while image gradient information serves as a driving force to align and refine segment edges according to gradient changes.

Compared to other models presented in Table~\ref{tableabk}, our proposed method demonstrates a much smaller variance on the Dice score. The standard deviation (s.t.d.) of the Dice score for Diff-mcs is 4.69, which is smaller than that of the other methods. The significant decrease in the variance indicates that Diff-mcs is a more stable and accurate model. The accurate prior probability distribution provided by the diffusion probabilistic model improves the level set segmentation results.  

\subsubsection{Segmentation Results on NIH Dataset}

\begin{table}
\caption{Comparison on the NIH Dataset}
\label{table2}
\setlength{\tabcolsep}{3pt}
\renewcommand{\arraystretch}{1.2}
\begin{tabular}{p{80pt}|p{50pt}|p{50pt}|p{50pt}}
\hline
\hspace{0.8cm}Method &
\hspace{0.3cm}maxDice & 
\hspace{0.3cm}minDice & 
\hspace{0.5cm}Dice\\
\hline
\hspace{0.1cm}Roth et al.\cite{roth2018spatial} & 
\hspace{0.5cm}88.96 &
\hspace{0.5cm}50.69 &  
\hspace{0.2cm}81.27$\pm$6.27  \\
\hspace{0.1cm}Zhou et al.\cite{zhou2017fixed} & 
\hspace{0.5cm}90.85 &
\hspace{0.5cm}62.43 &
\hspace{0.2cm}82.37$\pm$5.68 \\
\hspace{0.1cm}Cai et al.\cite{cai2019pancreas} & 
\hspace{0.5cm}90.85 &
\hspace{0.5cm}62.43 &  
\hspace{0.2cm}83.70$\pm$5.10  \\
\hspace{0.1cm}Li et al.\cite{li2019probability} & 
\hspace{0.5cm}91.08 &
\hspace{0.5cm}53.61 &  
\hspace{0.2cm}84.19$\pm$5.73  \\
\hspace{0.1cm}Yu et al.\cite{yu2018recurrent} & 
\hspace{0.5cm}91.02 &
\hspace{0.5cm}62.81 &  
\hspace{0.2cm}84.50$\pm$4.97  \\
\hspace{0.1cm}Hu et al.\cite{hu2020automatic} &
\hspace{0.5cm}91.64 &
\hspace{0.5cm}61.79 & 
\hspace{0.2cm}85.49$\pm$4.77  \\
\hspace{0.1cm}Chen et al.\cite{chen2022target} &
\hspace{0.5cm}92.28 &
\hspace{0.5cm}74.63 & 
\hspace{0.2cm}86.38$\pm$3.18  \\
\hspace{0.1cm}Tian et al.\cite{tian2023two} & 
\hspace{0.5cm}92.43 &
\hspace{0.5cm}74.13 &
\hspace{0.2cm}87.67$\pm$5.70 \\
\hspace{0.1cm}The proposed &
\hspace{0.5cm}\textbf{95.15}&  
\hspace{0.5cm}\textbf{75.70}&  
\hspace{0.2cm}\textbf{88.97$\pm$3.16}\\
\hline
\end{tabular}
\label{table2}
\end{table}

A similar result can be found in the NIH dataset, where Diff-mcs also demonstrate accurate and robust performance. Compared with the most recent methods as demonstrated in Table~\ref{table2}, Diff-mcs achieve the best Dice score of 88.97$\pm$3.16. Its s.t.d. decreased by an average of 2.02 compared to the previous eight methods, which indicates its stability. Additionally, our model achieves a maxDice score of 95.15, a minDice score of 75.70, and improves at least 1.3 average Dice scores relative to the second-ranked result.

As shown in Fig.~\ref{nih}, most of our segmentation results overlap with ground truth for the NIH dataset. In instances where the pancreatic volume and shape are relatively uniform, such as PANCREAS\_28 and PANCREAS\_72, there is a good overlap between predicted values and actual values. The second stage optimizes the inaccurate edges of the coarse segmentation. In PANCREAS\_74 and PANCREAS\_52, it can be observed that a large part of the green inaccurate edges are calibrated to the real edges.

\subsubsection{Segmentation Results on MSD Dataset}

As shown in the first subplot of Fig.~\ref{msd}, the coarse segmentation result contains three outliers, which are green tiny segmentation on the top-right of the pancreas. In the fine stage, the multi-cue level set method eliminates the outliers produced by the diffusion probabilistic model. In the 2nd and the 3rd subplots of Fig.~\ref{msd}, there are obvious excessive segmentations of the green area in the pancreatic tails, which are improved in the fine stage. 

The Dice scores on the MSD dataset can be found in Table \ref{table3}. Compared to the most recent methods, Diff-mcs achieve the Dice score of 87.18$\pm$4.78, which is 1.62 higher than the second-best method. The maxDice increased by 4.02 compared to \cite{zhang2021automatic}. Its s.t.d. is 0.19 lower than \cite{yu2018recurrent}, and the Dice score is 5.02 higher, which maintains a stable performance while improving the effect.

\subsection{Ablation Study}

\begin{table}
\caption{Comparison on the MSD Dataset}
\label{table3}
\setlength{\tabcolsep}{3pt}
\renewcommand{\arraystretch}{1.2}
\begin{tabular}{p{80pt}|p{50pt}|p{50pt}|p{50pt}}
\hline
\hspace{0.8cm}Method &
\hspace{0.3cm}maxDice & 
\hspace{0.3cm}minDice & 
\hspace{0.5cm}Dice\\
\hline
\hspace{0.1cm}Chen et al.\cite{chen2022pancreas} & 
\hspace{0.8cm}- &
\hspace{0.8cm}- & 
\hspace{0.2cm}76.60$\pm$7.30 \\
\hspace{0.1cm}Yu et al.\cite{yu2018recurrent} & 
\hspace{0.5cm}91.37 &
\hspace{0.5cm}66.53 & 
\hspace{0.2cm}82.16$\pm$4.97 \\
\hspace{0.1cm}Chen et al.\cite{chen2022target} & 
\hspace{0.5cm}91.87 &
\hspace{0.5cm}67.33 & 
\hspace{0.5cm}84.97 \\
\hspace{0.1cm}Zhang et al.\cite{zhang2021automatic} & 
\hspace{0.5cm}93.01 &
\hspace{0.5cm}69.26 & 
\hspace{0.5cm}85.56 \\
\hspace{0.1cm}The proposed & 
\hspace{0.5cm}\textbf{97.03} & 
\hspace{0.5cm}64.65 & 
\hspace{0.2cm}\textbf{87.18$\pm$4.78}  \\
\hline
\end{tabular}
\label{table3}
\end{table}

We conduct ablation studies on the multi-cue level set method, in which we specifically discuss the impact of the second stage model in scenarios including no rough position (Position), no texture cues(T), and no prior cues(P).
\subsubsection{Rough Position}
the accurate selection of the initial level set position is one of the important factors that affect the effectiveness of level set methods. In our model, the initial curve position is determined based on coarse segmentation results. In ablation studies, we set the initial contour as a circular curve with a fixed size in the middle of the original image. Table~\ref{table4} shows that inappropriate initial contours can lead to improper contraction or expansion during curve evolution when segmentation relies only on cues without prior contour information. Although our model remains effective, using the multi-cue level set method without rough position reduces accuracy.

\subsubsection{Texture Cues}
Compared to grayscale values, texture features have stronger stability against image noise and lighting variations. Therefore, texture cues can improve the robustness of the algorithm and reduce the impact of noise and other image changes on the segmentation results. We validate the influence of texture cues on the level set method by directly excluding the texture cues from our proposed model. Although the multi-cue level set method still works without texture cues, its average Dice score decreases by 0.26 compared to the complete model. Without texture cues, the s.t.d. increases from 4.00 to 6.00 which coincides with the fact that the texture cues help stabilize the segmentation. By analyzing the texture features of the pancreas, we can differentiate them more accurately from surrounding tissues.

\subsubsection{Prior Cues}
The prior probability map of coarse segmentation plays a crucial role in the entire model. Experimental observations indicate that without prior guidance, the model loses its original effectiveness. When the fitting term performs poorly, it fails to converge to the pancreas because organs other than the pancreas often stick together. Furthermore, removing texture cues along with the absence of prior cues leads to even worse results. The Dice decreased by 26.89 and 27.36 respectively, with minDice dropping as low as 21.7, indicating a loss of capability in the model. 

\begin{table}
\caption{The results of ablation studies}
\label{table4}
\setlength{\tabcolsep}{3pt}
\renewcommand{\arraystretch}{1.2}
\begin{tabular}{p{90pt}|p{40pt}|p{40pt}|p{60pt}}
\hline
\hspace{0.5cm}Ablation study&
\hspace{0.25cm}maxDice & 
\hspace{0.25cm}minDice & 
\hspace{0.5cm}Dice\\
\hline
\hspace{0.1cm}Position\&G\&T\&P  & 
\hspace{0.45cm}98.35 & 
\hspace{0.45cm}75.33 & 
\hspace{0.2cm}92.11$\pm$4.00  \\
\hspace{0.1cm}G\&T\&P& 
\hspace{0.45cm}98.00 &
\hspace{0.45cm}75.79 & 
\hspace{0.2cm}91.76$\pm$4.24  \\
\hspace{0.1cm}Position\&G\&P & 
\hspace{0.45cm}98.35 &
\hspace{0.45cm}58.28 & 
\hspace{0.2cm}91.85$\pm$6.00 \\
\hspace{0.1cm}Position\&G\&T& 
\hspace{0.45cm}91.45 &
\hspace{0.45cm}21.79 & 
\hspace{0.2cm}65.22$\pm$12.96  \\
\hspace{0.1cm}Position\&G & 
\hspace{0.45cm}92.04 &
\hspace{0.45cm}21.75 & 
\hspace{0.2cm}64.75$\pm$12.73  \\
\hline
\end{tabular}
\label{table4}
\end{table}
\subsection{Threshold Analysis}

\begin{figure}[ht!]
\centering
\includegraphics[width=3.5in]{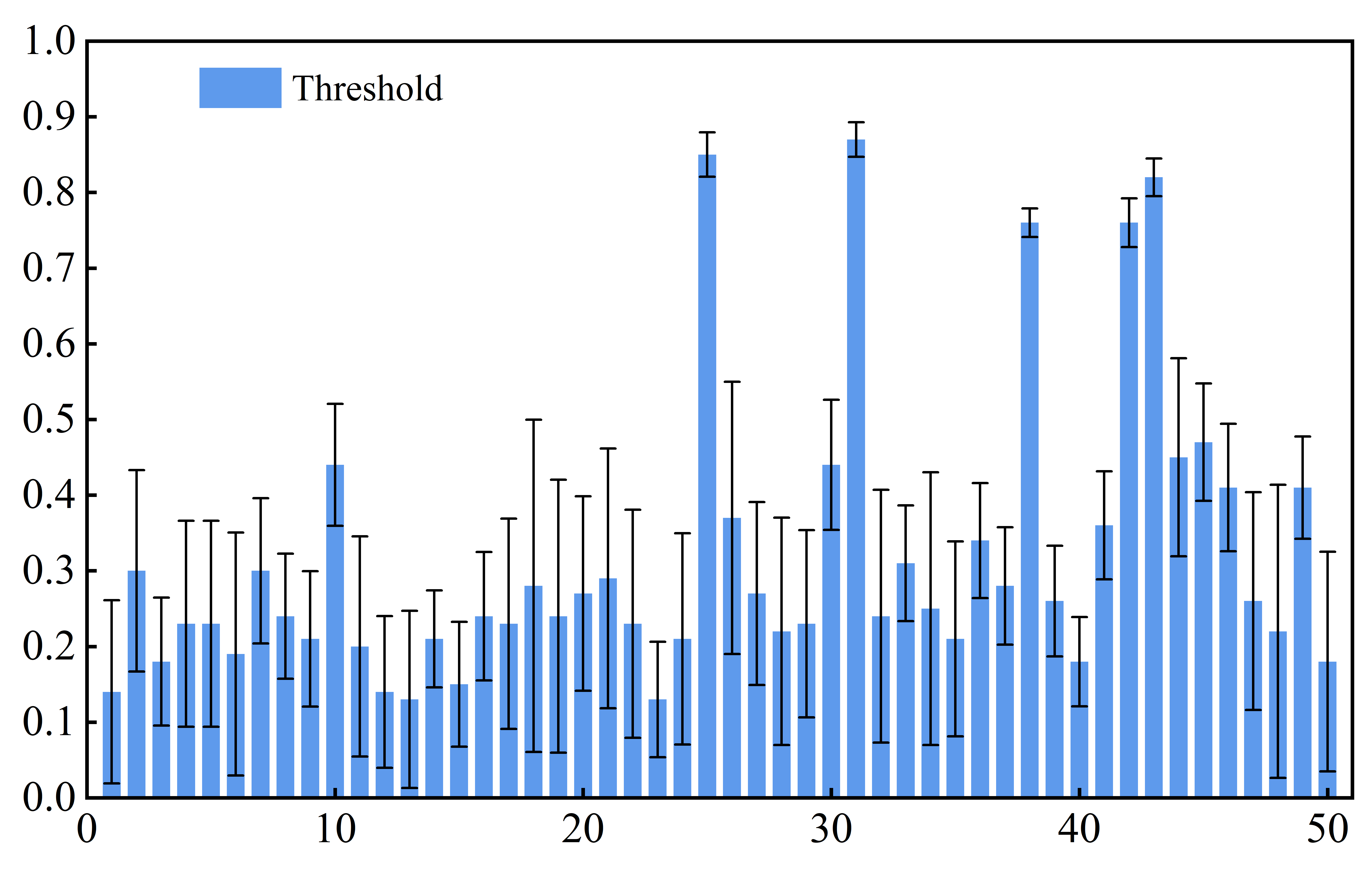}
\caption{Threshold distribution when different prior probability distributions achieve maximum values.}
\label{Threshold}
\end{figure}

As mentioned in the introduction, from the aspect of the diffusion probabilistic model, a constant threshold is not enough for accurate segmentation. In this section, we analyze the best thresholds for the prior probability maps. In Fig.~\ref{Threshold}, the bar chart represents the thresholds yielding maximum Dice scores for different instances, while the error bars indicate the variances in Dice scores at different thresholds for each instance. It can be observed that the threshold values achieve maximum Dice scores vary depending on the prior probability map, with instances having larger thresholds exhibiting smaller variances and greater stability. On the other hand, instances with smaller threshold values generally have higher variances, indicating a significant impact of threshold selection on results. Therefore, it is necessary to find an adaptive threshold or calibrate edges to determine edges for the diffusion probabilistic model segmentation, as it greatly influences segmentation outcomes for most cases.

\subsubsection{Comparative Analysis}
The images generated by GANs may lack details in the background and edge areas, resulting in a blurry appearance. This could be due to the challenges and instability during the training process of GANs. On the other hand, images generated by the diffusion probabilistic model prioritize detail preservation and image restoration, hence exhibiting relatively clearer details in the background section.

\begin{figure}[ht!]
\centering
\includegraphics[width=3.5in]{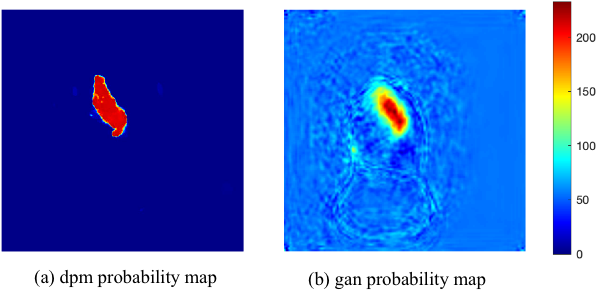}
\caption{The probability map is visualized, with the probability map from the diffusion probabilistic model on the left and the probability map from the GAN model on the right, both representing the same instance.}
\label{gan}
\end{figure}

\begin{table}
\caption{Comparative Analysis}
\label{table5}
\setlength{\tabcolsep}{3pt}
\renewcommand{\arraystretch}{1.2}
\begin{tabular}{p{110pt}|p{40pt}|p{40pt}|p{40pt}}
\hline
\hspace{1.5cm}Method &
\hspace{0.2cm}maxDice & 
\hspace{0.2cm}minDice & 
\hspace{0.5cm}Dice\\
\hline
\hspace{0.1cm}Diffusion Probabilistic Model & 
\hspace{0.4cm}97.98 &
\hspace{0.4cm}71.97 & 
\hspace{0cm}88.04$\pm$6.40 \\
\hspace{0.1cm}GAN(SegAN)\cite{xue2018segan} & 
\hspace{0.4cm}92.49 &
\hspace{0.4cm}44.38 & 
\hspace{0cm}78.83$\pm$8.54 \\
\hline
\end{tabular}
\label{table5}
\end{table}

As shown in Fig.~\ref{gan}, the segmentation results of the diffusion probabilistic model are only fuzzy near the edges, while the results of GANs segmentation are blurry throughout the background. Such extensive data fluctuations are not suitable as priors and may lead to additional errors. From Table ~\ref{table5}, it is evident that the diffusion probabilistic model performs better. During training, SegAN trains much faster than the diffusion probabilistic model but exhibits overfitting and significant performance differences in pancreas segmentation. There is a notable difference between the maximum and minimum Dice scores, particularly with GAN's minimum being too low.

\subsubsection{Uncertainty Analysis}

\begin{figure*}[ht!]
\centering
\includegraphics[width=5in]{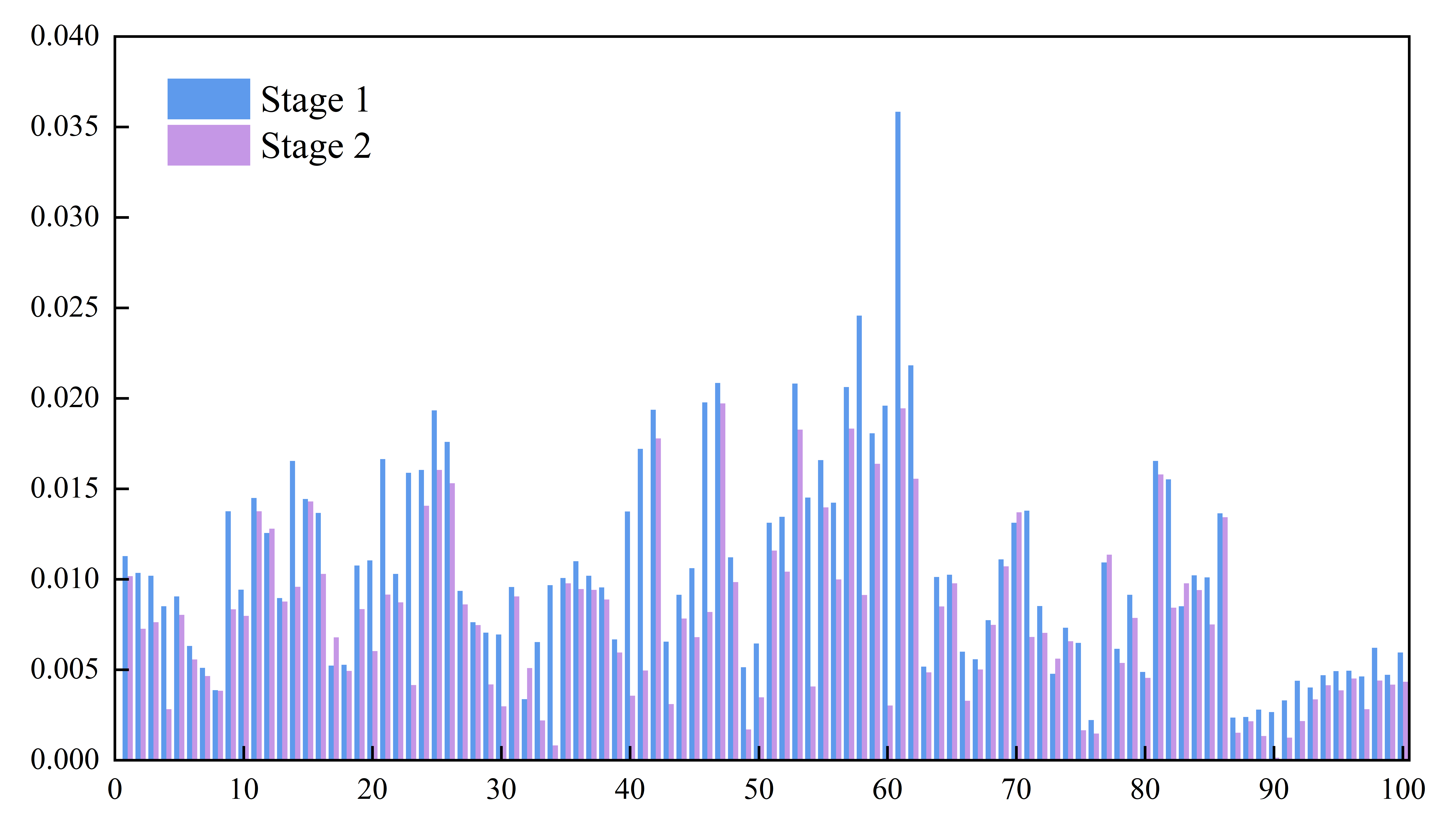}
\caption{In different instances, the variance of pixel points between the prior probability map based on the diffusion probabilistic model for segmentation and the resulting map of Diff-mcs. The pink color represents the variance of the prior probability map from the diffusion probabilistic model for segmentation, while the blue color represents the variance of the Diff-mcs result map.}
\label{var}
\end{figure*}

In this section, we compare the variances of the prior probability map and the result map obtained by Diff-mcs. As shown in Fig.~\ref{var}, the variance of Diff-mcs is generally lower than that of the prior probability map. This indicates that there is less variation in predicted results between neighboring pixels or regions, resulting in a smoother spatial variation in the segmentation result map. Furthermore, it also suggests that there is similar confidence in prediction among pixels at different locations, further validating the consistency and reliability of Diff-mcs.

\section{Conclusion}
In this study, we propose a segmentation model that combines the diffusion probabilistic model with the multi-cue level set method. Specifically, we use the prior probability map generated by the diffusion probabilistic model as the prior cues for the multi-cue level set method. This approach enhances image segmentation techniques by obtaining initial localization of the pancreas based on prior cues and guiding the level set method through a comprehensive feature vector composed of multiple cues. Our research findings indicate that due to limitations in the structure and generation process of the diffusion probabilistic model, it is difficult to find an adaptive threshold in the generated probability map to accurately describe edges. However, the multi-cue level set method effectively compensates for low confidence issues on coarse segmented edges. Furthermore, compared to other generative models, due to the stability and interpretability, when combined with the multi-cue level set method, the diffusion probabilistic model is more suitable as a prior. We validate Diff-mcs on three public datasets and achieve Dice scores of 92.30$\pm$4.69, 88.97$\pm$3.16, 87.18$\pm$6.78 respectively. Our method achieves more accurate and stable segmentation results compared to state-of-the-art approaches. We conduct ablation studies to verify the performance and importance of rough position provided by coarse segmentation, texture cues, and prior cues. Additionally, uncertainty analysis shows that our model’s performance after the multi-cue level set method processing is more stable than previous approaches. Our model has significant advantages in achieving high levels; although there may not be significant improvements in achieving lower scores which also provides directions for further research and improvement.

\section*{References}
\bibliographystyle{IEEEtran}
\bibliography{IEEEabrv,Bibliography}
\def\refname{\vadjust{\vspace*{-2.5em}}} 

\begin{IEEEbiography}[{\includegraphics[width=1in,height=1.25in,clip,keepaspectratio]{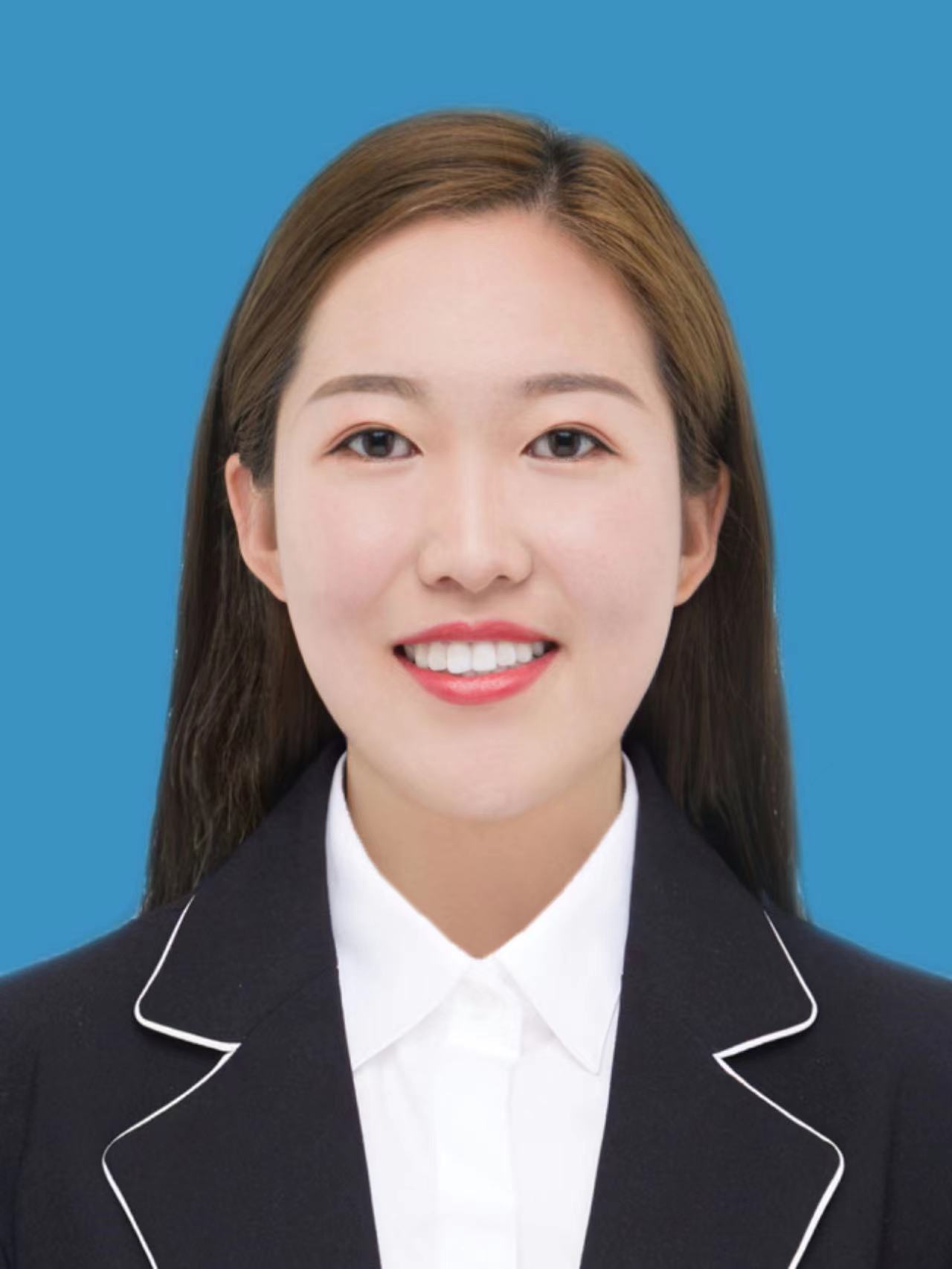}}]{Yue Gou,} received the bachelor’s degree from the Dalian Maritime University, Dalian, China, in 2021.     

She is currently pursuing the Ph.D. degree with the School of Mathematics, Harbin Institute of Technology.
Her research interests include image segmentation, partial differential equations, and deep learning.
\end{IEEEbiography}
\begin{IEEEbiography}
[{\includegraphics[width=1in,height=1.25in,clip,keepaspectratio]{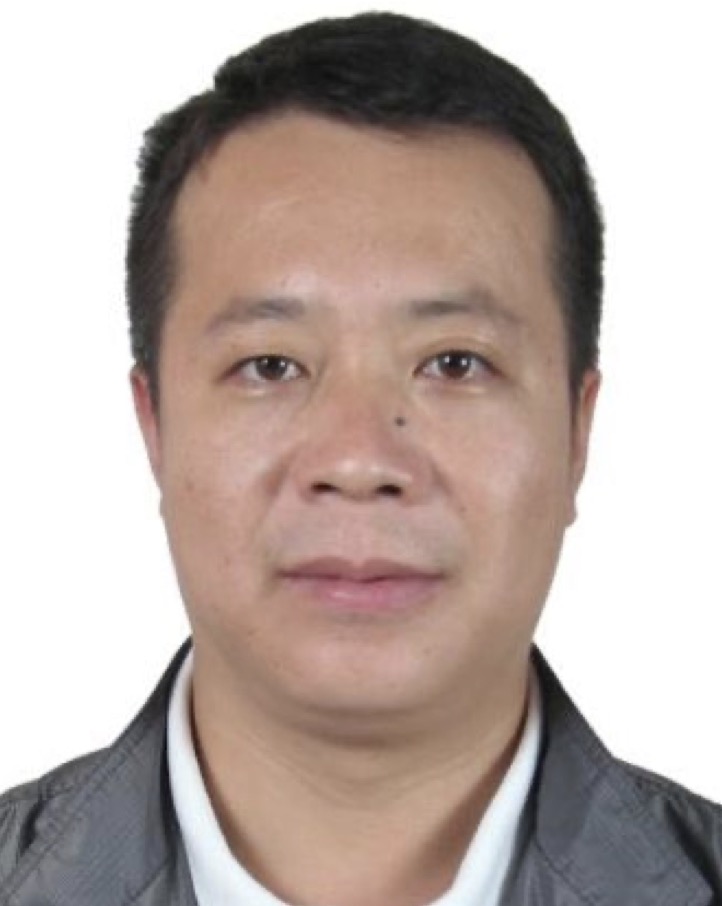}}]{Yuming Xing,} received the Ph.D. degree from the Harbin Institute of Technology, Harbin, China, in 2005.

He is currently a Professor with the School of Mathematics, Harbin Institute of Technology. His research interests include harmonic analysis, deep learning, partial differential equations, and image processing.
\end{IEEEbiography}
\begin{IEEEbiography}
[{\includegraphics[width=1in,height=1.25in,clip,keepaspectratio]{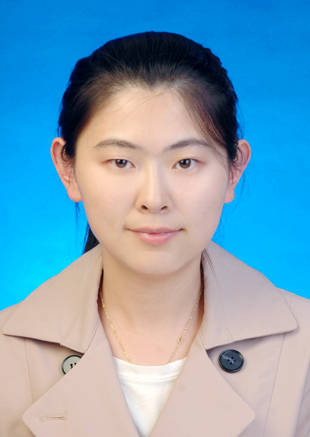}}]{Shengzhu Shi,} received the Ph.D. degree from Harbin Institute of Technology, Harbin, China, in 2022.

She is currently a lecturer in the School of Mathematics, Harbin Institute of Technology, Harbin, China. Her research interests include numerical methods for partial differential equations, uncertainty quantification, and deep learning.
\end{IEEEbiography}
\begin{IEEEbiography}
[{\includegraphics[width=1in,height=1.25in,clip,keepaspectratio]{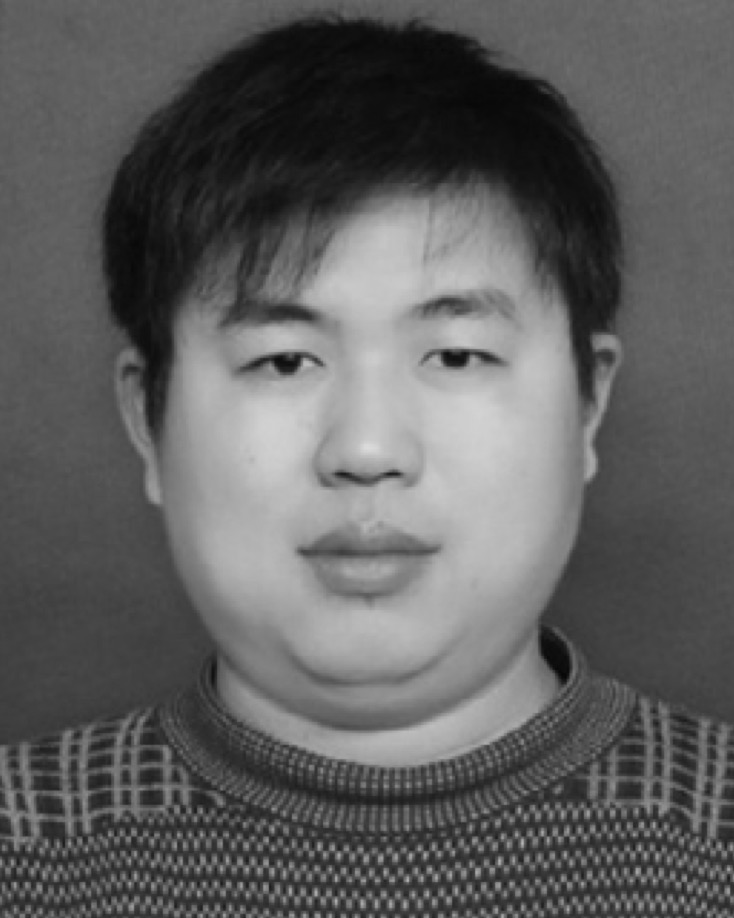}}]{Zhichang Guo.} received the Ph.D. degree from Jilin University, Changchun, China, in 2010.

He is currently a Professor with the School of Mathematics, Harbin Institute of Technology, Harbin, China. His research interests include partial differential equations, nonlinear analysis, mathematical methods in image analysis, and deep learning.
\end{IEEEbiography}
\end{document}